\documentstyle[12pt]{article}%
   \newcommand{\cent}[1] {\begin{center}#1\end{center}}
   \newcommand{\loo}{\,\raisebox{-.5ex}{$\stackrel{<}{\scriptstyle\sim}$}\,}
   \newcommand{\goo}{\,\raisebox{-.5ex}{$\stackrel{>}{\scriptstyle\sim}$}\,}
\begin{document}
\cent{\LARGE\bf
Fragmentation Phase Transition in Atomic Clusters II\\
--- Symmetry of Fission of Metal Clusters ---}
\cent{M.E. Madjet, P.A. Hervieux $^\diamond$, D.H.E. Gross, and O. Schapiro}
\cent{Hahn-Meitner-Institut
Berlin, Bereich Theoretische Physik,Glienickerstr.100 \\14109 Berlin, Germany\\
and\\
Freie Universit\"at Berlin\\
$^\diamond$ Institut de Physique, Lab LPMC,1 Bd Arago, F57078 Metz Cedex 3, France}
\cent{\today}
\begin{abstract}
We present a statistical fragmentation study of doubly charged alkali (Li, Na,
K) and antimony clusters.  The evaporation of one charged trimer is the most
dominant decay channel (asymmetric fission) at low excitation energies. 
For small sodium clusters this was quite early found 
in molecular dynamical calculations by Landman et al.
\cite{barnett91}. For
doubly charged lithium clusters, we predict Li$_{9}^{+}$ to be the preferential
dissociation channel.  As already seen experimentally a more symmetric fission
is found for doubly charged antimony clusters.  This different behavior
compared to the alkali metal clusters is in our model essentially due to a larger fissility
of antimony. This is checked by repeating the calculations for Na$_{52}^{++}$
with a bulk fissility parameter set artificially equal to the value of Sb.
\end{abstract}
\section{Introduction}
Doubly ionized clusters are not stable if the Coulomb repulsion energy between
the two positive holes exceeds the binding energy. The first experimental
evidence for Coulomb explosion of doubly charged clusters is due to Sattler et
al. in 1981 \cite{sattler81}. Cluster stability of multiply charged simple
metal clusters is usually studied after either increasing the internal energy
or the charge state by laser excitation \cite{naeher94a} or by collision with
charged particles \cite{chandezon95}. The size of the surviving clusters
depends on their charge state. The higher the charge the larger the cluster must
be to compensate the electrostatic pressure by adhesive forces. It is worth
noting that the critical sizes measured in these two experiments are different.
This might be due to different temperatures at which the clusters are produced
by the two methods. Also the deexcitation mechanism is different for different 
clusters.
For singly charged alkali metal clusters, experiments done
by Br\'{e}chignac et al. have shown that the excess of internal energy is
dissipated by successive evaporations of either neutral monomers or dimers
\cite{brechignac94b,brechignac89,brechignac90b}. In contrast, singly
charged antimony clusters relax their excess of energy by evaporation of
tetramers \cite{rayane89}.

Small doubly  charged alkali  clusters, due to the large surface tension of the
bulk material and due to the mobility of the electrons, fission asymmetrically.
Because of strong shell effects the emission of a charged trimer is the most
preferential dissociation channel.  This asymmetric fission of doubly charged
alkali clusters is in sharp contrast to fission of nuclei where the charge
degree of freedom is strongly linked to that of the mass by the symmetry force.
Therefore nuclear fission is much more symmetric. The experiments determined
the smallest mass a stable cluster can have with given charge, the critical
size of stability (clusters with mass smaller than 36 for $Z=2$ become
unstable)
\cite{brechignac94b,brechignac90,brechignac91}. It is, however, not true that
extreme asymmetric fission is a genuine feature of fission of doubly charged
metal clusters.  In a recent experiment Br\'{e}chignac et al.
\cite{brechignac95} found that larger doubly charged antimony clusters fission
the more symmetrically the larger the clusters are. Why does fission occur
symmetrically in some cases and asymmetrically in others ?.

The traditional method to describe the fission of atomic clusters is by
molecular dynamics ({\sl MD}). Here one follows the classical Newtonian
equation of motion of the interacting many-atom system. In contrast to
Lenard-Jones systems for which the interatomic potential is known, metal
clusters, due to the delocalization of the valence electrons, do not have well
defined interatomic potentials. Therefore, it is difficult to carry out {\sl
MD} calculations for metal clusters. One has to combine this with an explicit
treatment of the electronic degrees of freedom of the whole cluster like e.g.
a Kohn-Sham calculation \cite{barnett91} which is restricted to smaller
systems. By using a n-body potential (which reproduces the bulk
properties) to model neutral metal clusters, L\'{o}pez et al.\cite{lopez94}  have
investigated the fragmentation process at low excitation energies and for
clusters containing no more than 14 atoms by {\sl MD}.  Garcias et al. in a
series of papers \cite{Garcias95} used density-functional theory in two jellium
spheres to evaluate the fusion barriers for different doubly charged alkali
metal clusters.

Encouraged by the experience that the dynamics of interacting many-body systems
is often ergodic and is thus mainly controlled by the structure of the
accessible N-body phase space, we try Microcanonical Thermodynamics. A detailed
introduction into Microcanonical Thermodynamics is given in \cite{gross153}.

In this paper we investigate the coulombic fission of doubly charged metal
clusters of less than 60 atoms as a function of the excitation energy. We study
in particular how the excitation energy will be partitioned among the internal,
translational, rotational and charged degrees of freedom.  In using the
Microcanonical Metropolis Monte Carlo method ({\em MMMC}), described in details
in \cite{gross141,gross153}, see also the first paper of this series \cite{gross154}, we
present the fragmentation of doubly charged clusters of the elements Li, K, Na
and Sb. In the next paper of this series \cite{gross152} we will study the
implication of the accessible phase space on the fragmentation of multiply
($Z\ge2$) charged alkali clusters. A fourth paper \cite{gross157} will discuss
the relation of the fragmentation phase-transition to the liquid-gas transition
of the bulk.

\section{The effect of the bulk entropy}
The internal entropy of larger clusters is close to the entropy of the bulk
material at the same specific excitation energy. It is just one of the main
advantages of {\em MMMC} that it does not follow the individual motion of each
atom but allows incorporating known properties of the bulk material. This is
the more important as the internal behavior of the cluster becomes quite
complicated near structural transitions of the bulk. E.g. near melting high
anharmonicities evolve\cite{gross140} which are extremely difficult to describe
microscopically.

Knowledge of the specific heat at constant pressure of metals ( commonly at
atmospheric pressure) is fundamental for the description of their thermodynamic
behavior. Using this quantity and the latent specific heat, one can compute
other thermodynamic properties like the entropy or the internal energy as a
functions of temperature \cite{gross140}.

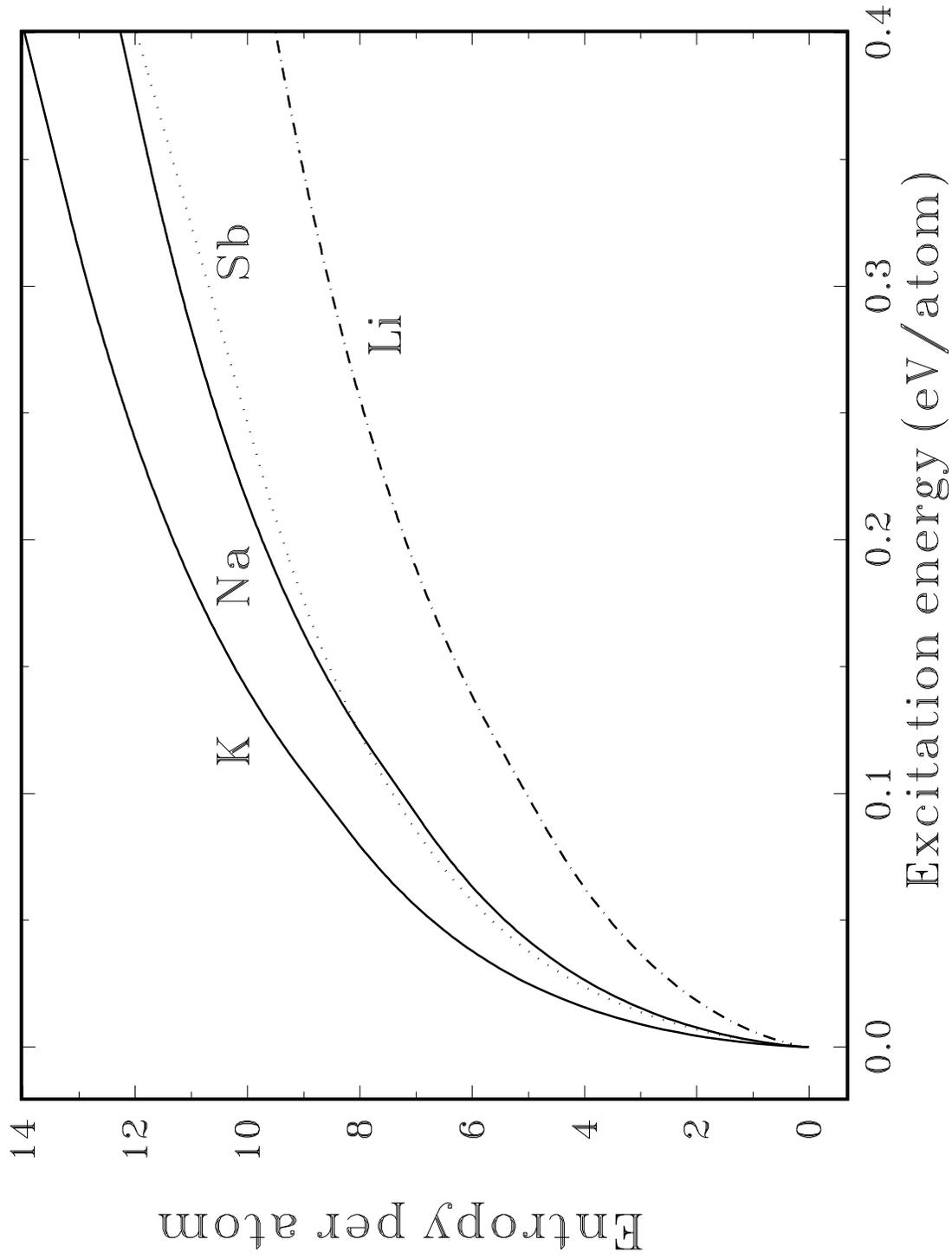
\begin{figure}
\caption{Specific entropy of bulk sodium, potassium, lithium and 
antimony at atmospheric pressure as a function of 
the specific internal energy.}
\label{fig1}
\end{figure}

Experiments involving alkali metal clusters are often carried out at
finite temperature, generally above the bulk melting temperature (for Na
$T_m\approx 400$ K). The clusters behave more like hot liquid droplets and the
harmonic (Debye) approximation is no longer valid to describe correctly the
internal behavior of the clusters.  For the particular case of antimony
clusters, however, which have a much higher bulk-melting temperature ($T_m=904$
K), one could expect the temperature of the clusters produced to be lower than
the melting temperature but still the effects of anharmonicities remain
important and have to be taken into account.  The experimental bulk specific
heat is taken from refs.\cite{alcock94,knacke91}. For antimony, as far as we
know, there are no data available for the very low temperatures below $T=298$ K.
Here anharmonicities become unimportant and the specific heat can be computed
within the Debye model, i.e. assuming the internal degrees of freedom to be
harmonic oscillators. The dependence of the specific entropy $s$ on the
specific excitation energy $\varepsilon$ for the different elements studied are
shown in Fig.\ref{fig1}.  The Debye temperature gives the temperature above
which all vibrational modes are excited. Below this temperature, the modes
begin to be frozen out. For alkali metals the element with the smallest Debye
temperature has the largest entropy over the entire range of excitation
energies (the Debye temperatures $\theta_D$ are reported in Tab.I).  This is a
consequence of the dependence of the specific heat at low temperature on the
Debye temperature.  One remarks also that the specific entropy of antimony is
very close to that of sodium.  In fact, these two elements have similar Debye
temperatures.

As pointed out by N. Ju and A. Bulgac \cite{ju93}, the finite temperature
properties of  simple metal clusters, in particular phase transitions, are
dominated by the ionic degrees of freedom  as a consequence of the very small
contribution to the entropy coming from the electrons.  Therefore, in the
present model we ignored the influence of the valence electrons on the entropy
(e.g. through shell effects as discussed in ref.\cite{frauendorf95}).

Our physical scenario of statistical cluster fission is as follows: We assume
the dynamics of a cluster to be sufficiently chaotic after it was excited by a
laser or by a passing heavy ion. If this hypothesis is correct the cluster will
fission highly ergodically. That means many replicas of the reaction with the
same macroscopic initial conditions, impact parameter, excitation energy etc.,
explore the details of the topological structure of the N-body phase space
which is accessible under the constraints of the global conservation of energy,
angular momentum, mass, and charge. We further {\em assume} that the stochastic
coupling of the fragment motion has a short range very similar to the strong
and short range friction which acts between the moving fragments in nuclear
fragmentation
\cite{gross16}. This hypothesis is still to be proven.  However, it allows to
simplify highly the general complicated expansion dynamics of the decay. By
this assumption it is possible to subsume all details under two fit parameters,
the freeze-out volume outside of which the stochastic coupling of the fragments
seizes and the maximum internal excitation energy $\varepsilon_{max}$ of the
fragments.

Inside the freeze-out volume with a radius $$R_{sys}=r_f*N^{1/3}$$ the cluster
and its fragments are strongly stochastically coupled and explore the
accessible phase space. Outside of it the various fragments disassemble
independently (besides an eventual Coulomb interaction between the fragments
).

The second parameter of our fragmentation model as defined in 
ref.\cite{gross141} is the maximum specific internal energy per atom which is
allowed for the fragments ($\varepsilon_{max}$ see Tab.I). In analogy to our
previous study on the fragmentation of hot sodium clusters, $\varepsilon_{max}$
is estimated by the assumption that inside the freeze-out configuration  the
excited clusters should live longer than the lifetime of the freeze-out
configuration itself. Fragments excited to higher energies decay inside the
freeze-out configurations and these excitations are counted as excitations of
the daughter-fragments. Moreover, the system is assumed to be in equilibrium
within the freeze-out volume. Therefore, the evaporation times of the fragments
(calculated within the Weisskopf model) corresponding to $\varepsilon
\le \varepsilon_{max}$ must be also larger than the characteristic time
associated to the interaction between the atoms (period of vibration) inside
each fragment.

It was mentioned by N. Ju and A. Bulgac \cite{ju93} the melting and boiling 
temperatures of sodium clusters do not seem to be strongly dependent on the
particle number.  This behavior is at variance to that of tin clusters
\cite{lai96} and justifies our assumption to take $\varepsilon_{max}$ (at least
for Na) to be independent of the particle number.

\section{Alkali Metal Clusters}
The ground state binding energies are computed within the metallic liquid drop
model. First we need the barriers for charged particle decay. These must be
calculated from the electrostatic energy between two charged metallic
spheres. This is done by iterating the induced image charges.  According to a
recent paper of Seidel and Perdew \cite{seidel94}, who have solved the
classical image potential paradox, we have modified the formula (13) given in
ref.\cite{gross141}. The parameters which enter in this formula are given in
Tab.I. In order to take into account the electronic shell effects which are
important for small clusters, we have used, if available, the experimental
binding energies instead of those from the metallic liquid drop model. For
singly charged clusters we have used the binding energies given in
refs.\cite{brechignac94b,brechignac89,brechignac90b} for Li, Na, and K,
respectively. From the knowledge of the experimental ionization potentials
\cite{dugourd92,kappes88,kappes86,heer87} it is possible to get the binding
energies for neutral clusters up to the size 26, 21, and 26 for Li, Na, and K,
respectively.\\ In our fragmentation model, since the properties of diatomic or
the triatomic molecules are well known, we used for dimers and trimers the
values of the vibrational frequencies and the calculated principal moments of
inertia given in Tab.I.  We have improved our model by taking polarization
effects into account \cite{gross152}. As discussed in details in ref.
\cite{gross152}, we do not allow the fragments to be closer than a certain
distance which is fixed at 1.0 \AA \hspace{0.1cm} for all the elements studied.

Initial clusters having 30, 42 and 52 atoms decay quite differently because of
shell effects in the daughters. For the sizes 30 and 42, the daughters from the
fission process may have both magic number of electrons (i.e.
$X_{30}^{++}\rightarrow X_{21}^{+}+X_9^{+}$ and $X_{42}^{++} \rightarrow
X_{21}^{+} + X_{21}^{+}$). Fission of Na$^{++}_{52}$ which does not fulfill the last
property was taken as a third example. We calculated the distribution of the
three largest average masses as a function of the specific excitation energy
for the different initial cluster sizes. As neutral monomers are not considered
as fragments they are not included in the mass distribution presented here.
The fragments can be either charged or neutral.

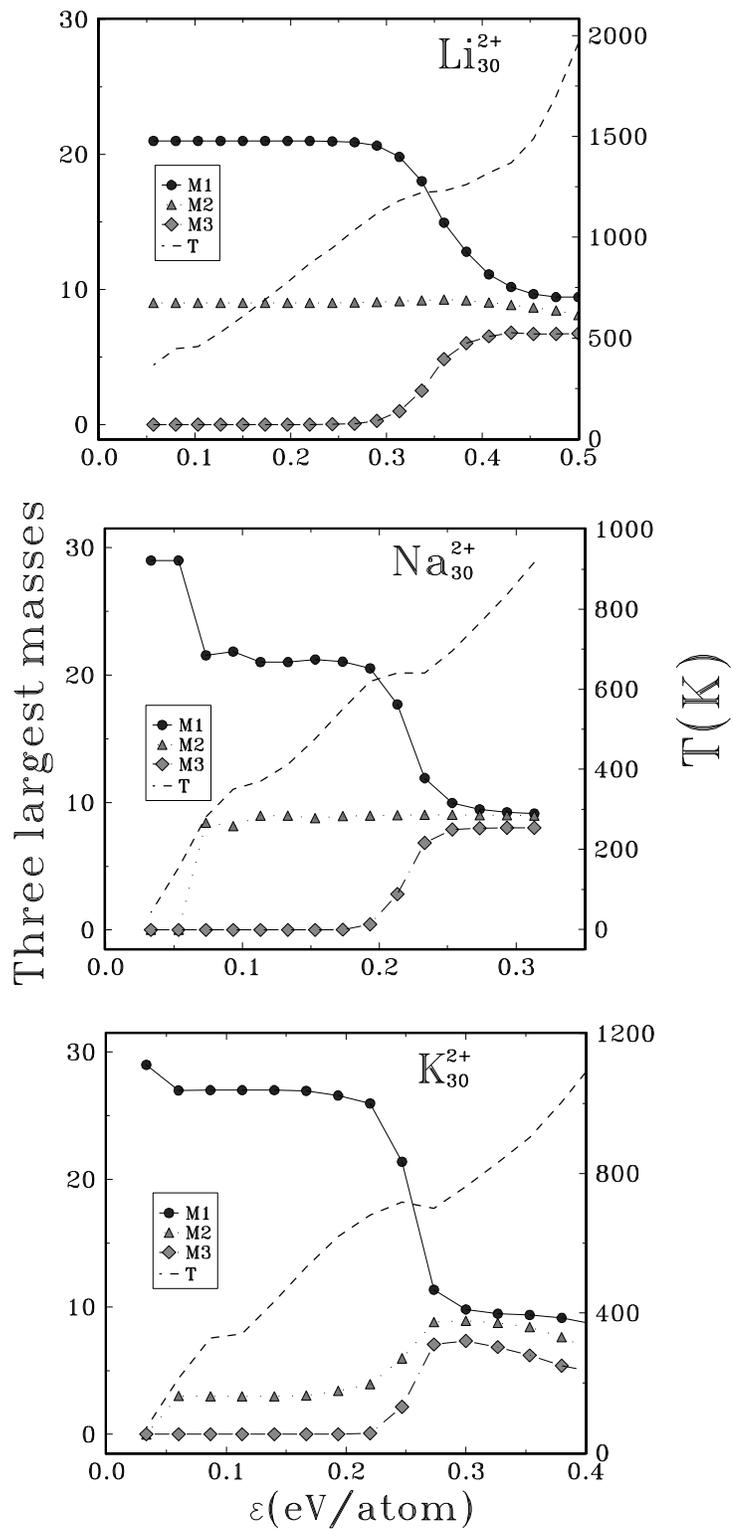
\begin{figure}
\caption{Average masses of the three largest fragments 
as a function of the specific 
internal energy for the clusters:(a) Li$_{30}^{2+}$, (b) Na$_{30}^{2+}$ and 
(c) K$_{30}^{2+}$.
The dashed curve shows the caloric curve $T(E)$ in Kelvin. In
this subsection quadrimers and heavier fragments are assumed to be conducting
spheres.}
\label{fig2}
\end{figure}

In Fig.\ref{fig2} we present the results for a fissioning cluster of size $30$.
Lithium clusters, at $\varepsilon\loo 0.25$ eV/atom decay only into
Li$_{21}^{+}$ and Li$_{9}^{+}$. One notes that this fission channel is the only
one. Above $\varepsilon\approx$ 0.3 eV/atom one sees the appearance of a
neutral fragment and at $\varepsilon\goo$ 0.4 eV/atom the system decays into
more than three fragments.  Within the statistical model the thermodynamic
temperature is in the energy range $0.05$ eV/atom $\le\varepsilon\le<0.4$ eV/atom a
linear function of $\varepsilon$ varying from $335$ K to $1200$ K.
In contrast to Na and K for which the singly
charged trimer is the most stable cluster, Li$_{9}^+$ has the highest
stability.  We have checked that at low excitation energy ($\varepsilon<0.25$
eV/atom) the fission pattern is not sensitive to the internal entropy of the
fragments. In this case the sampling of the internal excitations was performed
by using a harmonic level density (classical Einstein model) \cite{gross141}
with a single vibrational frequency. Our finding of a dominating decay into a
Li$_{9}^+$ is in contradiction with the experiment \cite{brechignac94}. In
fact, experimentally the trimer is still the preferential fission channel, the
production of Li$_9^+$ exists but with a much lower probability. As already
mentioned above, one cannot explain the experimental results in using only
energetic considerations and it seems that also our statistical approach is not
able to reproduce the experimental trends for the fission of Li$^{++}_n$. One
must recall that the statistical theory ignores any information about the
dynamics of fragmentation.  Consequently, one may think that there is a
dynamical mechanism which plays a role in the fission process. Other possible
reasons for the failure of our statistical multifragmentation model here may be
the neglect of nonsphericities of the Li trimers which taken into account could
well lower the barriers for the trimer decay. At this moment we are unable to
decide this.

The calculations for Na$_{30}^{2+}$ are presented in fig.\ref{fig2}b. 
At very low energy $\varepsilon < 0.07$ eV/atom the system evaporates one
neutral monomer. Above this energy the system decays into one large and one
small singly charged fragment.  There is a competition between the emission of
Na$_{3}^+$ and Na$_{9}^+$ the latter fragment having the largest probability.
One notes that besides these two channels we have also the presence of other
fragments but with much lower probability.  At an energy above $0.20$ eV/atom,
the system prefers to fragment into three or more fragments of intermediate
sizes. It is worth noting that the energy at which the
multifragmentation takes place is higher for Li than for Na and this is a
direct consequence of a larger cut-off of the bulk excitation energy
$\varepsilon_{max}$ of Li (see Tab.I).  

K$_{30}^{2+}$ decays, at very low energy, $\varepsilon < 0.07$ eV/atom, like
Na$_{30}^{2+}$ by evaporating one neutral monomer, see figure \ref{fig2}c. However,
above this energy the ejection of a singly charged trimer is the preferred
fission channel. When the excitation energy increases the number of channels
increases but trimer ejection remains the dominant mode of decay. For higher
energies ($\varepsilon > 0.2$ eV/atom) the system follows the same scenario as
described above for Na. Due to similar values of $\varepsilon_{max}$ for Na and
K, the multifragmentation mode appears at about the same energy.

Let us now discuss the clusters of sizes $42$, see fig.\ref{fig4} for Li,
below $0.1$ eV/atom we have emission of Li$_{9}^{+}$.  Above this energy, the
fragment Li$_{9}^+$ remains the most probable product of fission but one sees
the appearance of symmetric fission through the production of two fragments of
equal sizes.  There is also the presence of other fragments but with much less
probability.  When the energy is increasing the fission process becomes more
and more symmetric. This is clearly illustrated in Fig.\ref{fig4} where we have
plotted the relative probability as a function of the mass of the fragments at
two excitation energies.

For Na$_{42}^{++}$, below $\varepsilon \approx 0.1$ eV/atom we have evaporation
of one neutral monomer. For $0.1$ eV/atom$ < \varepsilon < 0.3 $ eV/atom there
is a competition between Na$_3^+$ and Na$_9^+$ and no presence of Na$_{21}^+$.
From 0.1 eV/atom to 0.2 eV/atom Na$_3^+$ is dominant. From 0.2 to 0.3 eV/atom
Na$_9^+$ is dominant.  Note, that compared to Na$_{30}^{++}$ the fragmentation
region shifts upwards in energy and becomes more sharp.  The thermodynamic
temperature shows at $\varepsilon\sim 0.3$ eV/atom a backbending signalizing a
phase transition of first order. Evidently this cluster is large enough to show
the fragmentation phase transition.
 
For K$_{42}^{++}$, the distribution is rather similar to that of K$_{30}^{++}$.
Here also the fragmentation is shifted to higher energy.  Again a first order
transition towards fragmentation is clearly seen as a backbending in
$T(\varepsilon)$.

The size $52$ is interesting to study since the fission into two fragments
having both closed shells is no longer possible. As it can be seen in
Fig.\ref{fig5}a, for Li, over the whole range of energy up to the transition
energy ($\varepsilon \approx 0.4$ eV/atom) Li$_9^+$ is the preferential
dissociation channel. In contrast with Li$_{42}^{++}$, the production of
Li$_{21}^{+}$ is non-existent.  For Na$_{52}^{++}$ and K$_{52}^{++}$, the
results are shown in Figs.\ref{fig5}b and c.  In all clusters of this size the
first order transition towards fragmentation can clearly be seen. A comment may
be necessary on the high thermodynamic temperatures the system can get in the
range of excitation energies studied: They are finally {\em larger} than the
boiling temperature of the bulk. Here one has to keep in mind that because of
the assumed short range of the possible interfragment friction our
calculations are at constant freeze-out {\em volume} not at constant pressure
as in studies of boiling of the bulk, see further discussion of this point in
the fourth paper of this series \cite{gross157}.

\section{Antimony Clusters}
It has been observed experimentally \cite{brechignac95} that antimony clusters
exhibit a completely different behavior with respect to alkali metal
clusters. For sizes around $44$, antimony prefers to fission into two singly
charged clusters with similar sizes and this can not be explained in terms of
electronic shell effects.

Due to the lack of experimental and theoretical data for this type of clusters,
we were forced to make several assumptions. We considered antimony
clusters to have metallic character and therefore the binding energies of the
charged species will be computed within the metallic liquid drop model.  Thus
the later finding of a  more symmetrical fission than Na-clusters (as found in
the experiments) is within our phase space model {\em not} due to a different
conductivity and charge distribution. Even with the extreme assumption of a
metallic conductivity as in alkali clusters Sb$^{++}$-clusters fission more
symmetrically.  For neutral and singly charged dimers and trimers we have used
the experimental binding energies given in ref.\cite{cabaud73} and their
moments of inertia were computed from the theoretical data of
\cite{sundararajan95}.  The surface energy $a_s$ is calculated from the bulk
value of the surface tension $\alpha$ ($a_s=4\pi r_s^2 \alpha$), here $r_s$ is
the Wigner-Seitz radius.

The computed mass distributions are shown in Fig.\ref{fig6} for the mass $30$,
$42$, and $52$.  At low energy, the system breaks into two singly charged
fragments having comparable sizes. As the energy is increasing one observes the
appearance of a third fragment namely a neutral dimer or a neutral quadrimer.
Depending on the initial cluster size, above a certain energy the preferential
third fragment is Sb$_4$. This production of Sb$_4$ is artificial since we have
taken the experimental binding energies for the dimers and trimers and used the
liquid drop model to compute the binding energies of the other fragments.  The
caloric curve $T(\varepsilon)$ is much more smooth than for Na-clusters. This
is so because this is far below the bulk-boiling of antimony which is at
$1907$ K.

Quantitative predictions by our statistical model of the symmetrical character
of the fission process for clusters having at least $52$ atoms compare
favorably with the experimental results \cite{brechignac95}.

The ground state fissility parameter $f$ is 
defined as the ratio of the Coulomb energy $E_C$ 
to the surface energy $E_S$ and within the metallic liquid drop model can be 
expressed as :

\begin{eqnarray}
f&=&\frac{E_C}{2E_S}=\chi\frac{Z^2}{N}\label{fiss1}\\
\chi&=&\frac{e^2}{4 r_s a_s}  
\label{fissi}
\end{eqnarray} 
where $Z$ and $N$ are the charge and the mass of the initial cluster.
Let us take the example of a cluster of size $52$ and charge $2$, $f=
0.12,0.13, 0.11$ and 0.62 for Li, Na, K and Sb respectively. 
One immediately notes that Sb has a fissility $\approx 5$ times larger 
than the other alkali metal elements.
 
Another useful quantity is the asymmetry parameter $\eta$ which is defined as
the ratio between the size difference of the two singly charged fragments to
the parent size. $\eta$ is a function of the temperature and of the initial
cluster size. If one wants to compare theoretical and experimental results, one
needs to have an estimate of the temperature at which the experiment was
carried out. It comes out from the model that for Sb$_{52}^{++}$, $\eta
\approx 0.15$ at a thermodynamic temperature of about $300$ K (far below 
the melting temperature $T_m=904$ K !). This must be compared to the
experimental result $0.27 \leq \eta \leq 0.34$ \cite{brechignac95}.
Unfortunately, in the ref.\cite{brechignac95} the temperature was not given.

A possible explanation of this disagreement is given by the fact that we have
used in our calculations the value of the bulk surface tension to compute the
surface energy $a_s^B$. However, we know that $a_s$ must be modified in order
to take finite size effects into account and usually the corrected value of
$a_s$ is larger than the one of the bulk (e.g. for Na, $a_s^B=0.80$ eV and
$a_s=1.02$ eV \cite{brechignac94b}).  Consequently according to formula
\ref{fissi}, the fissility decreases and the fission becomes less symmetrical.
In using the same ratio ($a_s/a_s^B$) as for Na, we have performed again the
calculations and we have found $\eta \approx 0.27$ which is in better agreement
with the experimental values.  Also non perfect metallic conductivity  would
lead to more symmetric fission.
 
Using the liquid drop model we observe that the fragmentation does not evolve
with increasing cluster size from an asymmetric to a symmetric fission which is
in disagreement with the experiment. A possible reason is : For low masses
(here $30$ and $42$), one observes experimentally an asymmetric fission with a
predominance for Sb$_5^+$ and Sb$_7^+$ as small fragments. This behavior seems
analogous to the shell effects of alkali metal clusters. In contrast to our
precedent study on alkali metals, and due to lack of experimental data no shell
effects have been included in our present calculations.

To gain a basic understanding of the role played by the surface energy on the
fission process, we have calculated the average three largest masses as a
function of the excitation energy for Na$_{52}^{++}$ in using a value of $a_s$
which corresponds to the same fissility as for Sb$_{52}^{++}$ (of course
in the pure liquid drop approximation for the binding energies).  The results
of the computation are shown in Fig.\ref{fig7}. One should mention that both Na
and Sb have nearly identical entropies over the entire excitation energy range
(see Fig.\ref{fig1}). The distribution is surprisingly similar to that of
Sb$_{52}^{++}$ (see Fig.\ref{fig6}c).  One concludes that the symmetrical
character of the fission seems to be mainly governed by the smaller surface
tension and the smaller Wigner-Seitz radius of bulk antimony in the same
combination of the two as is expressed by the fissility parameter $\chi$. There
is however, a remarkable difference to nuclear fission: In metal cluster
fragmentation one can vary charge and mass independently of one another within
some margins. Increasing the charge the fissility rises according to formula
(\ref{fiss1}). Instead of shifting to more symmetric binary fission,
Na$^{Z+}_n$ decays into several (up to Z) charged fragments. Below the
fragmentation transition one is heavy, often doubly charged, and  the others
are singly charged nine-mers or trimers. The heavy one has a low fissility
again. This is not possible in nuclear fission as there the mass to charge
ratio of the fragments can not vary so much.

Further, we hope that the experimentalists in the near future will be able to
measure the binding energies of singly charged antimony clusters. From these
values we could get a more realistic value of $a_s$. An indication of the
temperature at which the experiments are carried out is necessary.

\section {Conclusions}

We have shown that the symmetry of fission of doubly charged metal clusters is
ruled by the number of exit channels (phase space) which are energetically
accessible (calculated within the framework of our statistical fragmentation
model {\em MMMC}) and depends on:
\begin{itemize}
\item the initial cluster size,
\item the initial excitation energy,
\item the bulk surface tension and the bulk density (combined to the fissility
parameter $\chi$),
\item
and the electronic shell effects through the ground state binding energies of
the fragments.
\end{itemize} 

For Li the dominant decay channel is always  Li$_9^+$.  This behavior is in
contradiction to the experiment.  This disagreement might be due to a dynamical
mechanism or due to deformation effects  which are not taken into account in
our statistical model.  For Li$_{42}^{++}$ we find at high enough excitation
energies a symmetric fission (with a probability of $\approx 30\%$) as a
consequence of the closed electronic shell of Li$_{21}^+$.

At low excitation energy, below the multifragmentation transition, for Na and K
clusters, the fission is always asymmetric whatever is the initial size. For a
given size, when the excitation energy increases, the number of dissociation
channels increases as well.  There is a competition between the different
channels which are energetically favored by electronic shell effects.

This study demonstrates that the symmetry of the fission process which has been
observed experimentally for doubly charged antimony clusters of intermediate
size can be interpreted in terms of the larger bulk fissility parameter $\chi$.

Finally, our calculations have shown that it is possible to classify the
effects shaping the fission pattern into two types according to their origin :
electronic shell effects through the ground state binding energies and the
bulk fissility.
\section{Acknowledgment}

The Sonderforschungsbereich SFB 337 of the DFG supported this work by granting
a post-doc position (M.Madjet).
The Fachbereich Physik of the Freie Universtit\"at Berlin made this work
possible by giving access to their computer system.
\clearpage

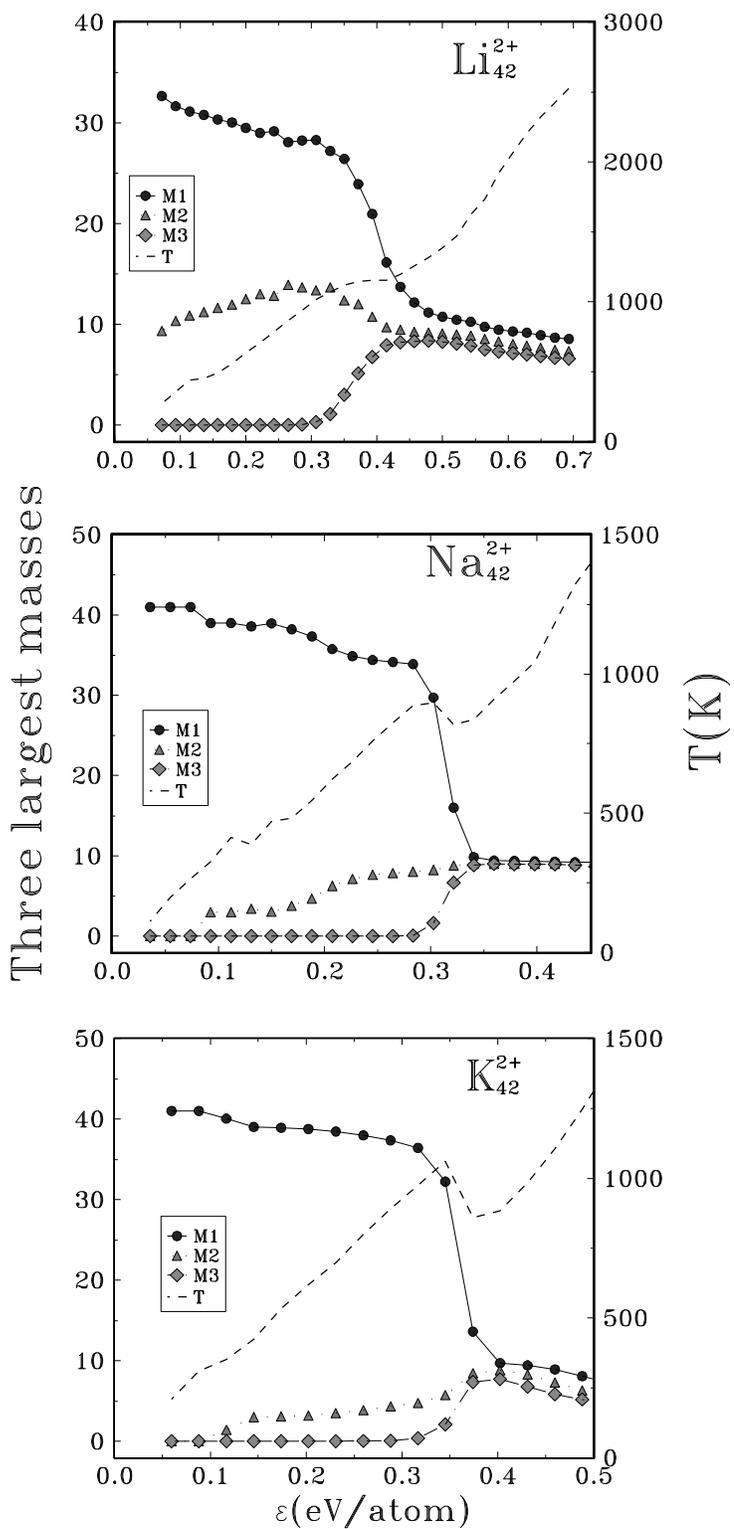
\begin{figure}
\caption{Same as Fig.2 but for: 
(a) Li$_{42}^{2+}$, (b) Na$_{42}^{2+}$ and 
(c) K$_{42}^{2+}$. }
\label{fig3}
\end{figure}

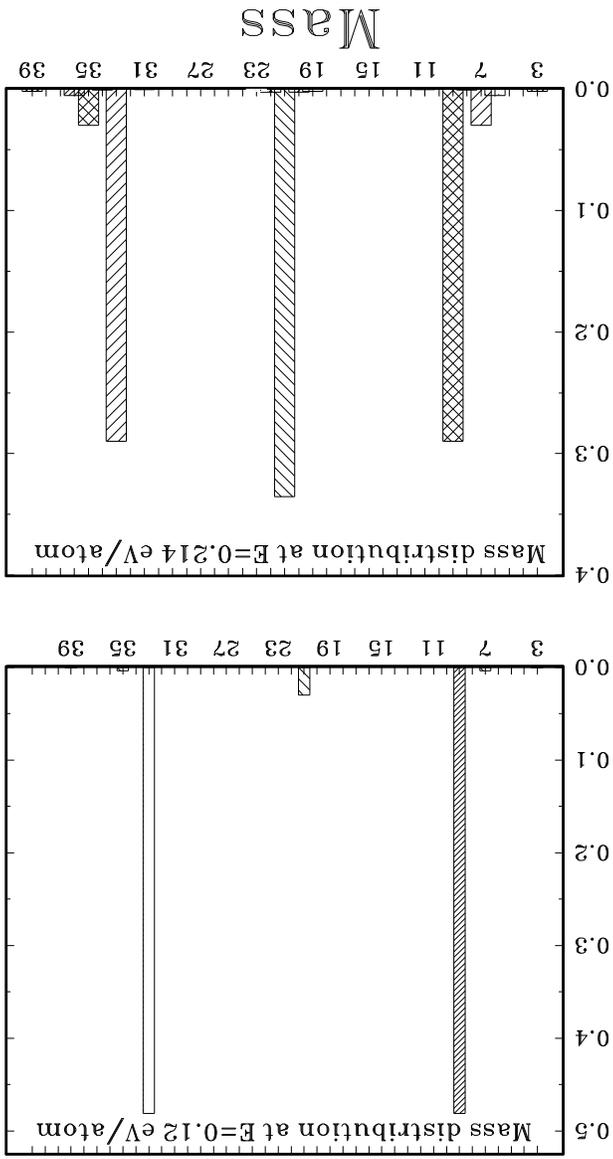
\begin{figure}
\caption{Mass distribution of Li$_{42}^{2+}$ at a specific excitation 
energy: 
(a) $\varepsilon=0.12$ eV/atom and (b) $\varepsilon=0.214$ eV/atom.}
\label{fig4}
\end{figure}

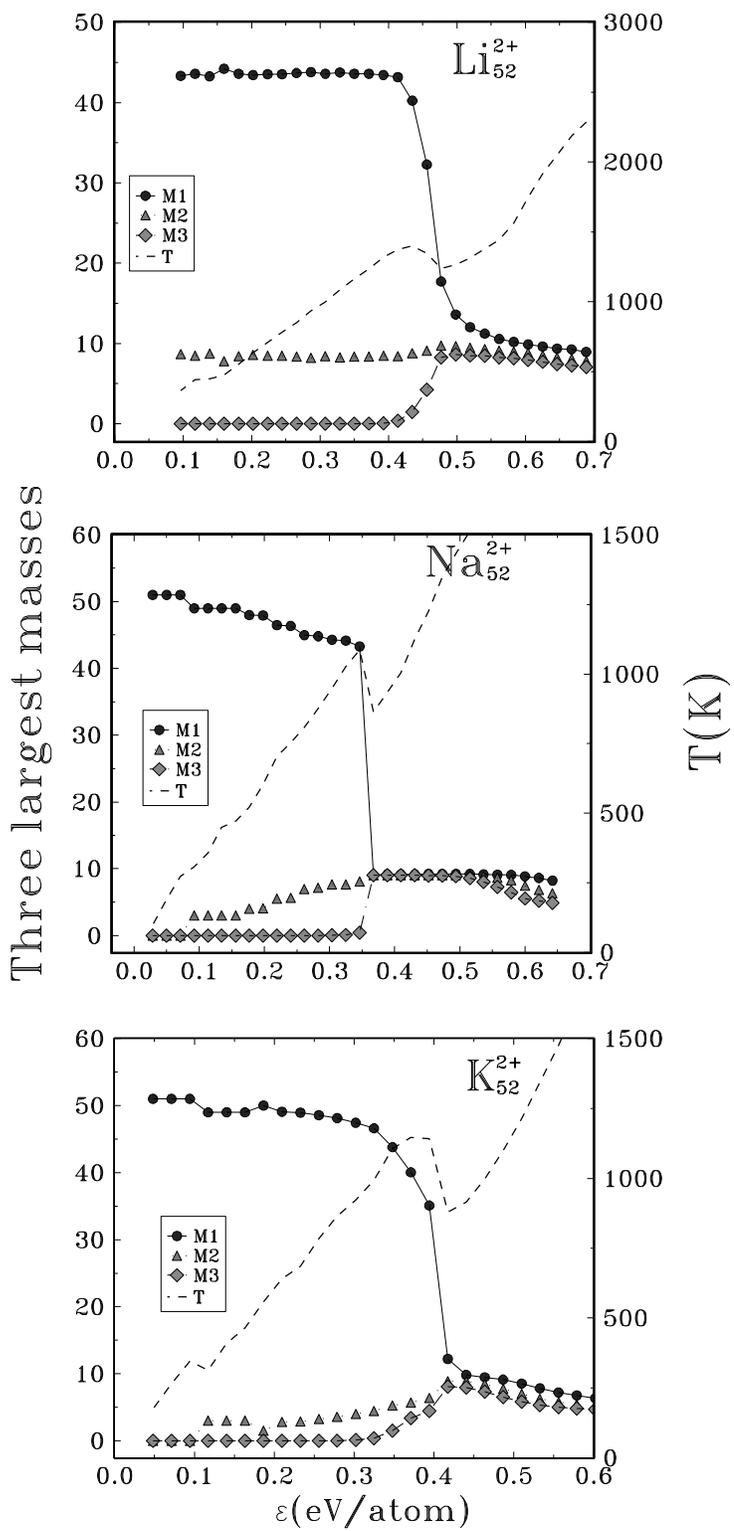
\begin{figure}
\caption{Same as Fig.2 but for: 
(a) Li$_{52}^{2+}$, (b) Na$_{52}^{2+}$ and 
(c) K$_{52}^{2+}$. }
\label{fig5}
\end{figure}

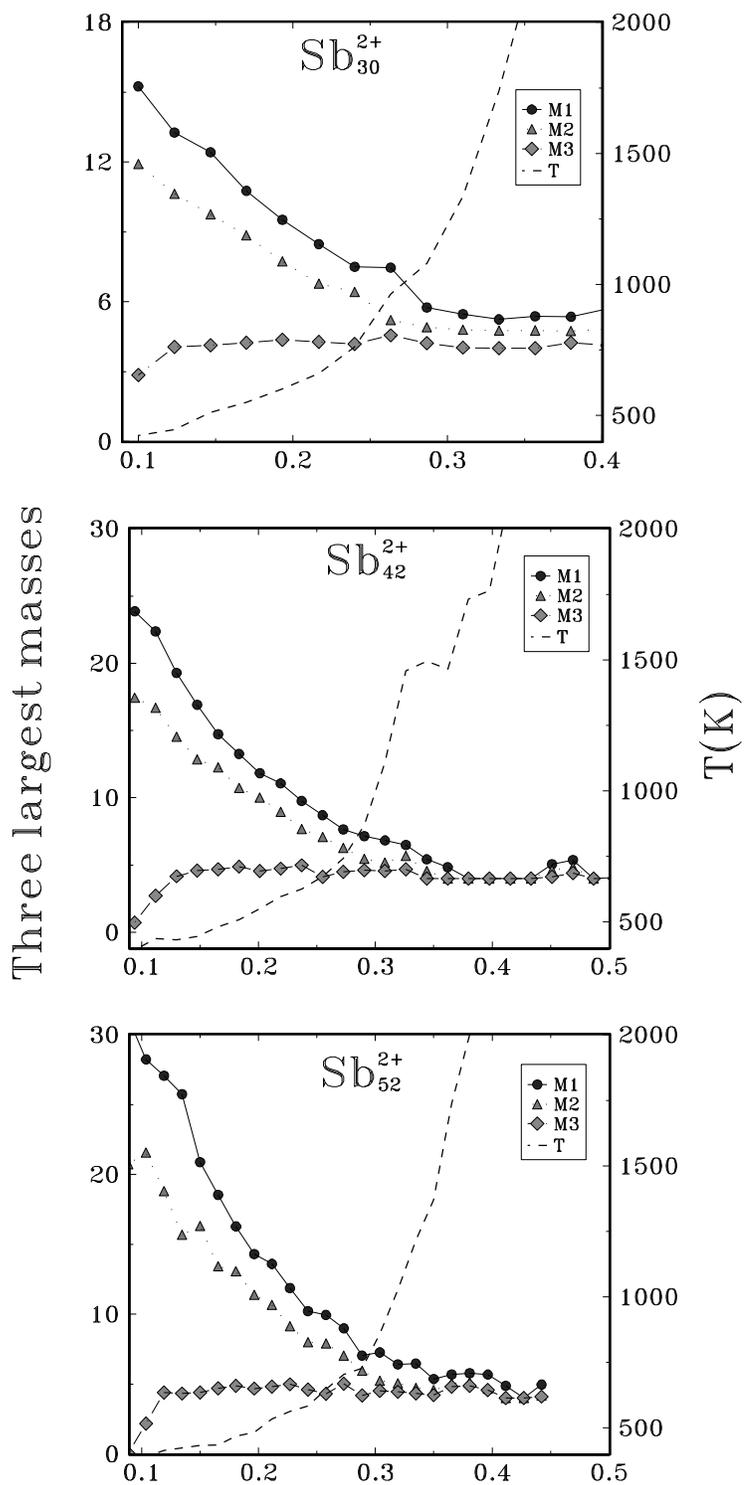
\begin{figure}
\caption{Same as Fig.2 but for: 
(a) Sb$_{30}^{2+}$, (b) Sb$_{42}^{2+}$ and 
(c) Sb$_{52}^{2+}$. }
\label{fig6}
\end{figure}

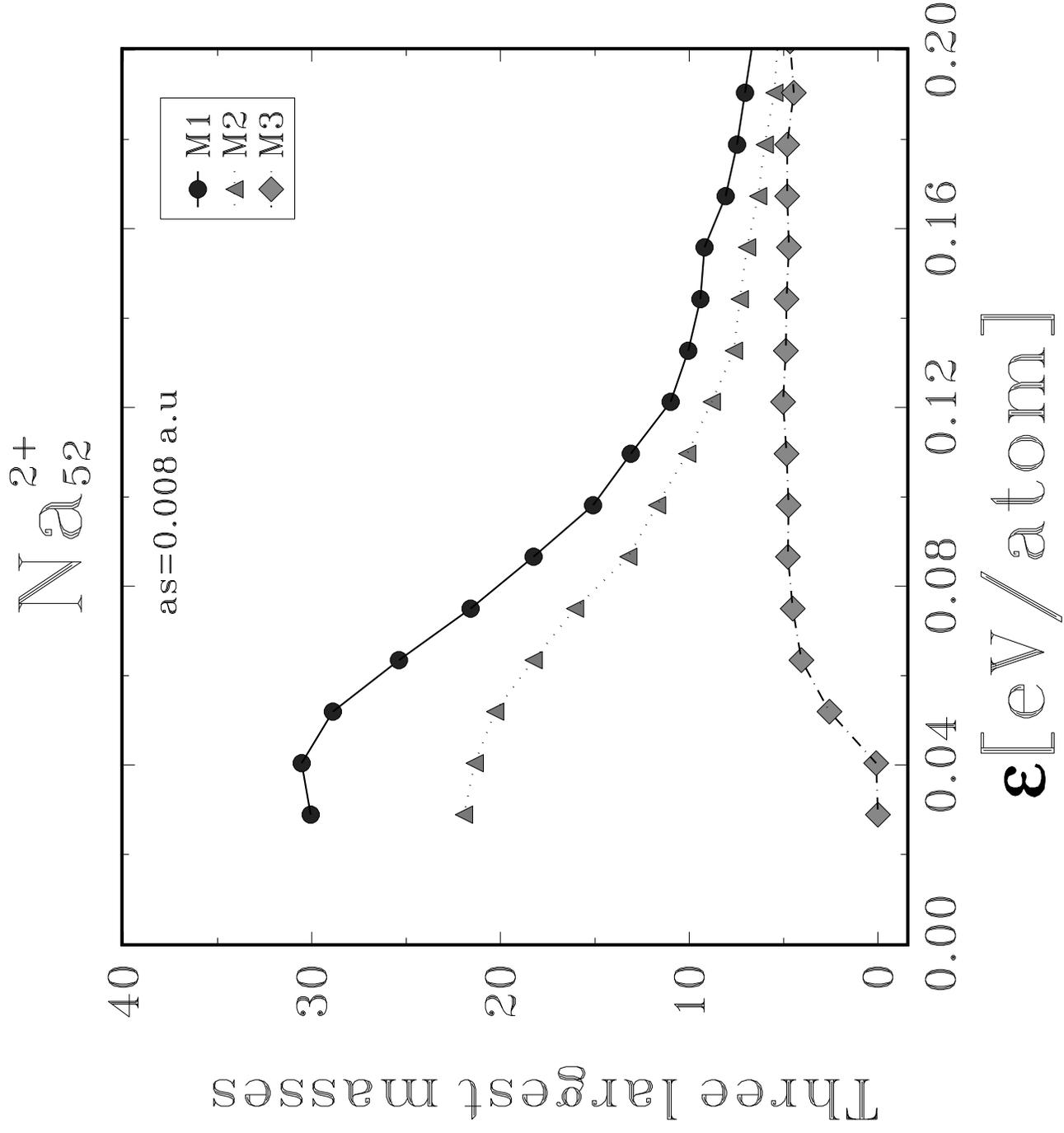
\begin{figure}
\caption{Average masses of the three largest fragments 
as a function of the specific 
internal energy for the cluster Na$_{52}^{2+}$  with a modified 
surface energy $a_{s}=0.218$ eV.}
\label{fig7}
\end{figure}

\begin{table}
\caption{Experimental and theoretical values of the different parameters used
in our calculations.  $r_s$ is the Wigner Seitz radius
\protect\cite{weast77}
$a_v$ is the volume cohesion energy and $a_s$ its surface part
\protect\cite{brechignac94b,weast77}. 
The work function for ionisation of the bulk $W_{\infty}$
\protect\cite{weast77,michaelson77}, the 
ionization energy of the atom $IP$ \protect\cite{weast77}, the radius of the
freeze-out configuration $r_f$, the principal moments of inertia for the dimer
$I_2$ \protect\cite{wang90,huber79} and  for the trimer $I_3$
\protect\cite{boustani87,poteau93,flad83,sundararajan95}, the Debye temperature
$\theta_D$ \protect\cite{burns,alcock94}, the melting and boiling temperatures
$T_m$ and $T_v$ \protect\cite{alcock94,weast77}, the dimer and trimer
frequencies $\omega_d$ \protect\cite{huber79} and $\omega_t$
\protect\cite{carlier91},the maximum specific internal energy of the bulk
$\varepsilon_{max}$, and the bulk fissility coefficient $\chi$. For simplicity
we used in all cases for $\varepsilon_{max}$ the boiling energy at normal
pressure. This corresponds to $\tau_{evap}=\tau_{\varepsilon_{max}}\approx
10^{-11}$secs. In the case of Sb the excitation of the fragments never reached
values of $\varepsilon = \varepsilon_{max}$} ~\\~
\newpage
\cent{
\begin{tabular}{|l|c|c|c|c|}
\hline
Element          &  Li    & Na      & K       & Sb   \\ 
$r_{s}$(\AA)     & 1.719    & 2.070     & 2.571     & 1.130   \\ 
$W_{\infty}$(eV) & 2.490    & 2.751     & 2.299     & 4.550   \\ 
$IP$(eV)         & 5.361    & 5.140     & 4.340     & 8.640   \\ 
$a_{s}$(eV)      & 1.301    & 1.020     & 0.980     & 0.397   \\ 
$a_{v}$(eV)      & 1.521    & 1.121     & 0.939     & 2.751   \\ 
$\theta_{D}$(K)  & 345.     & 150.      & 90.       & 140.    \\ 
$\omega_{d}$(eV) &$ 4.352E-2$ &$1.490E-2$ &$1.141E-2$ &$1.206E-2$ \\ 
$\omega_{t}$(eV)&$ 2.973E-2$ &$1.048E-2$ &$7.755E-3$ &$1.206E-2$ \\ 
$\varepsilon_{max}$(eV/atom) &0.460  &0.350   &0.330      &0.750  \\ 
$\tau(\varepsilon_{max})$(secs)&$1.85E-11$&$5.33E-11$&$3.15E-11$&$2.18E-8$ \\
$I_{2}/m_{0}$(\AA$^2$)    &3.56     &4.74      &7.64      &2.71  \\ 
$I_{3x}/m_{0}$(\AA$^2$)    &2.94     &4.84      &7.45      &3.15  \\ 
$I_{3y}/m_{0}$(\AA$^2$)   &7.88     &5.66     &8.63      &3.71  \\ 
$I_{3z}/m_{0}$(\AA$^2$)  &10.82     &10.49      &15.62      &6.86  \\ 
$T_{m}$(K)     &453.65     &370.95      &336.60      &904.00  \\ 
$T_{v}$(K)     &1600.00     &1156.00      &1033.00      &1907.00  \\ 
$r_{f}$ (\AA)         &3.20     &3.85      &4.78      &2.10  \\ 
$\chi$            &1.60  &1.69     &1.42      &7.96 \\
\hline
\end{tabular}}
\label{table1}
\end{table}
\clearpage

\end{document}